\documentclass[preprint,12pt,authoryear]{elsarticle}

\newcommand{\mnras}{MNRAS}
\newcommand{\aap}{AAP}
\newcommand{\icarus}{Icarus}
\newcommand{\araa}{ARAA}
\newcommand{\apj}{ApJ}

\newcommand{\aj}{AJ}
\newcommand{\pasp}{PASP}
\newcommand{\nat}{Nature}
\newcommand{\planss}{PLANSS}
\newcommand{\rmxaa}{RMXAA}

\usepackage{amssymb}
\usepackage{lineno}

\usepackage{ecrc}

\volume{00}

\firstpage{1}

\journalname{Icarus}

\jid{Icarus}

\jnltitlelogo{Icarus}

\journal{Icarus}

\begin{document}

\begin{frontmatter}



\title{Evidence for Pebbles in Comets}


\author[label1]{K.~A.~Kretke}
\address[label1]{Southwest Research Institute, 1050 Walnut Ave, Suite 300, Boulder, CO, 80302, USA.}
\author[label1]{H.~F.~Levison}

\begin{abstract}

	When the EPOXI spacecraft flew by Comet 103P/Hartley 2, it observed large
	particles floating around the comet nucleus.  These particles are likely
	low-density, centimeter- to decimeter-sized clumps of ice and dust.  While
	the origin of these objects remains somewhat mysterious, it is possible
	that they are giving us important information about the earliest stages of
	our Solar System's formation. Recent advancements in planet formation
	theory suggest that planetesimals (or cometestimals) may grow directly from
	the gravitational collapse of aerodynamically concentrated small particles,
	often referred to as ``pebbles.'' Here we show that the particles observed
	in the coma of 103P are consistent with the sizes of pebbles expected to
	efficiently form planetesimals in the region that this comet likely formed,
	while smaller pebbles are may be expected in the majority of comets, whose
	chemistry is often indicative of formation in the colder, outer regions of
	the protoplanetary disk.
	
\end{abstract}


\end{frontmatter}
%\linenumbers
\section{Introduction}
\label{sec:intro}

When the Epoxi spacecraft flew by Comet 103P/Hartley 2 it observed large
particles floating around the comet's nucleus \citep{AHearn.etal.2011}.  Based
on best estimates of the albedo of these objects, they appear to have a rather
steep size distribution in which the largest particles are thought to be 
$\sim$ 10 cm in size \citep{Kelley.etal.2013,Harmon.etal.2004}.  These particles
are larger than has been generally detected in the coma of other comets.  While
it is possible that the size of these particles is determined simply by the
erosive process inside the comet \citep{Meech.Svoren.2004}, another intriguing
possibility is that these particles may be left over relics of the formation
process.

In the past it has generally been assumed that the icy bodies in the outer
solar system were all built up in a hierarchical fashion.  In this picture,
km-sized comets are built from binary collisions of smaller bodies, while the
larger Kuiper belt objects and even the cores of the giant planets were formed
by accretion of these small ``cometesimals''
\citep{Stern.1996,Stern.Colwell.1997,Kenyon.2002,Kenyon.etal.2008}.  However,
there are important challenges in forming comets via this mechanism.  First, it
may not be possible to grow to binary collisions up to km-sized objects.  As
particles grow their sticking efficiency is reduced, possibly leading to a
regime in which collisions are more likely to lead to ``bouncing'' rather than
accretion \citep{Zsom.etal.2010,Guttler.etal.2010}.  Additionally, as small
micron sized particles grow into macroscopic mm to m-sized particles they
attain larger relative velocities, increasing the likelihood that collisions
will be destructive \citep{Blum.Wurm.2000,Blum.Wurm.2008}.  And while
mass-transfer precesses may allow lucky particles to grow beyond these
barriers, the timescale for formation of planetesimals becomes long enough to
become problematic for subsequent planet formation \citep{Windmark.etal.2012}
Additionally, even if these barriers can be overcome and planetesimals can be
formed, to grow the larger Kuiper belt objects in the time alloted by this
process the disk must have been extremely dynamically cold and 100 to 1000
times more massive than it is today \citep{Kenyon.etal.2008}.

However, more recently there have been theoretical and observational reasons to
suggest that instead of forming in this bottom up manner, planetesimals may
form directly by the gravitational collapse of a dense cloud of small particle
embedded in the protoplanetary disk (the gravitational instability or GI
hypothesis).  In particular, recent breakthroughs in theory and computer
simulation have demonstrated that under reasonable conditions particles can be
concentrated to the point that they will gravitationally collapse, as suggested
by the GI hypothesis \citep[see reviews
by][]{Chiang.Youdin.2010,Johansen.etal.2014}.  In these models particles that
are small enough to have their orbits strongly perturbed by aerodynamic drag,
yet large enough to still be decoupled from the gas (mm to m sized objects
depending on the gas disk properties) can self-clump, forming gravitationally
bound objects of 10-1000 km in size
\citep[e.g.][]{Johansen.etal.2007,Johansen.etal.2015}.

Furthermore, new observational signatures may support the idea that
planetesimals form via GI rather than through hierarchical
growth. For example, the size distribution of planetesimals in the asteroid
belt appears to be inconsistent with the predictions of classical planetesimal
collisions, and instead may be showing us that the larger asteroids had to
directly form as relatively large objects (\citet{Morbidelli.etal.2009}, although see \citet{Weidenschilling.2011} for an alternative interpretation).
Additionally, the delicate wide binaries in the Kuiper Belt are unlikely to be
made by normal processes such as collisions or binary exchanges, instead they
are most easily made by a collapsing, fragmenting gravitationally bound pebble
``cloud'' \citep{Nesvorny.etal.2010,Parker.etal.2011}. 
Furthermore, if one combines the observed activity of comets with the strength the dusty surface layers of comets, the dust on cometary surfaces must be in the form of relatively large particles in order for water sublimation to power cometary activity \citep{Skorov.Blum.2012,Blum.etal.2014}.  All of various lines of evidence combined paint a consistent picture in which small icy bodies in the outer-Solar system formed from collapsing clouds of small ``pebbles''.

If Kuiper belt objects did indeed form from the gravitational collapse of
clouds of small pebbles, there may be signatures of these formative pebbles
extant in comets today.  While some pebbles may destructively fragment during
the planetesimal formation process, so long as the initially formed
planetesimal was under 100 km in radius it is expected that most of the pebbles
are unlikely to collide at high enough speed to cause fragmentation
\citep{Jansson.Johansen.2014}.  Additionally, while thermal processing will
alter the cometary surface, creating lag deposits that mask the underlying
structure of the comet, only the surface layers of the comet likely have been
significantly altered by thermal processing \citep{Mumma.etal.1993}.  This may
leave pristine material beneath the lag layer.  With this in mind, in this
\emph{Note} we are interested in investigating the speculative proposition that
the large particles observed in the coma of Hartley 2 may be remnants of the
initial pebbles which formed the comet.  To this end we will address the issue
of the expected size of pebbles in different regions of the protoplanetary disk
and the possibility of predictable trends.

In this paper we look at the large particles in the coma of comet Hartley 2 and
show that their sizes are consistent with the size of pebbles expected to
efficiently form planetesimals.  In section \ref{sec:model} we describe from a
theoretical standpoint how large pebbles need to be in protoplanetary disks to
be concentrated by well-understood processes.  In section \ref{sec:comets} we
use the chemical composition of comet Hartley 2 to estimate its formation
location so that we can place the sizes of the particles in it coma in context
with the theoretical expectations.  Finally, in section \ref{sec:conclusions}
we summarize our results and discuss the implications and what future data may
help further elucidate the formation mechanism of comets.

\section{Theoretical Expectations of Pebble Sizes} \label{sec:model}

There are three main ideas for how pebbles can be concentrated to the degree
necessary for gravitational collapse in protoplanetary disks, turbulent eddies,
pressure bumps and vortices, and the streaming instability \citep[see][for a
review]{Johansen.etal.2014}.  Each of these processes will only work on
particles within a limited range of sizes, determined by the particles' Stokes
numbers ($\tau$).  The Stokes number is defined as $\tau \equiv t_s \Omega$,
where $t_s$ is the stopping time of the particle due to aerodynamic drag and
$\Omega \equiv \sqrt{G M_*/r^3}$ is the Keplerian orbital frequency around a
star of mass $M_*$ at a heliocentric distance $r$. The stopping time is 
\begin{equation} 
  t_s = 
   \left\{
      \begin{array}{ll}
		  \frac{a \rho_s}{c_s \rho_g},& {\rm if}~a < \frac{3}{2}\lambda\\
		  \frac{2 a^2 \rho_s}{3 c_s \rho_g}, & {\rm otherwise}
	  \end{array}
   \right.
\end{equation}
\citep{Adachi.etal.1976} where $\lambda$ is the gas mean-free-path, $c_s$ is
the sound-speed, $\rho_s$ is the density of the solid particle, and $\rho_g$ is
the local gas density.  

Turbulent eddies should concentrate particles whose stopping times are
comparable to the eddy turnover times, which corresponds roughly to $\tau\sim
10^{-5}-10^{-4}$ in typical protoplantary disks
\citep[e.g.][]{Cuzzi.etal.2008,Cuzzi.etal.2010}, however \citet{Pan.etal.2011}
found that strong clustering of particles in this size range may be too
difficult to form the number of required planetesimals.  Particle concentration
in pressure bumps
\citep[e.g.]{Whipple.1972,Haghighipour.Boss.2003,Kretke.etal.2009}, and vortices
\citep[e.g.]{Barge.Sommeria.1995,Lyra.etal.2008} is most effective for
particles with $\tau\sim 1$.  The streaming instability
\citep{Youdin.Goodman.2005,Johansen.etal.2007}  was shown to effectively
concentrate particles with $\tau$ in the range of between $10^{-2}$ and $1$
\citep{Bai.Stone.2010,Carrera.etal.2015}.  As the streaming instability is a
linear instability that has been robustly shown to function under physically
reasonable conditions, for the remainder of this paper we will take the
streaming instability range as the size scale of interest.

To determine the sizes of particles susceptible to the streaming instability,
we must know the midplane gas density and temperature profiles in
protoplanetary disks.  Unfortunately, these parameters cannot currently be
measured directly in the regions where comets are thought to have been formed
($\sim$5 to $\sim$ 50 AU), and theoretically depend upon parameters that are
expected to vary over the disk lifetime (such as the mass-accretion rate
through the disk) and/or are generally poorly constrained (such as the grain
opacity and the disk viscosity).  However, we can construct a reasonable
fiducial model based upon existing constraints and discuss how our results are
sensitive to our assumptions.  

For the protoplanetary disk structure we assume a disk heated
by a combination of viscous heating and stellar irradiation.  We use the models
of \citet{Bitsch.etal.2015}, which utilize a full radiative transfer model to
calculate the 2D thermal structure of an axisymmetric protoplanetary disk.
This model has the advantage of more accurately calculating the temperature
profiles than more standard 1+1d or 2 layer calculations
\citep[e.g.][]{Chiang.etal.2001,DAlessio.etal.2005,Kretke.Lin.2010}, particularly in regions that may be self-shielding.

\begin{figure}
	\includegraphics[width=\textwidth]{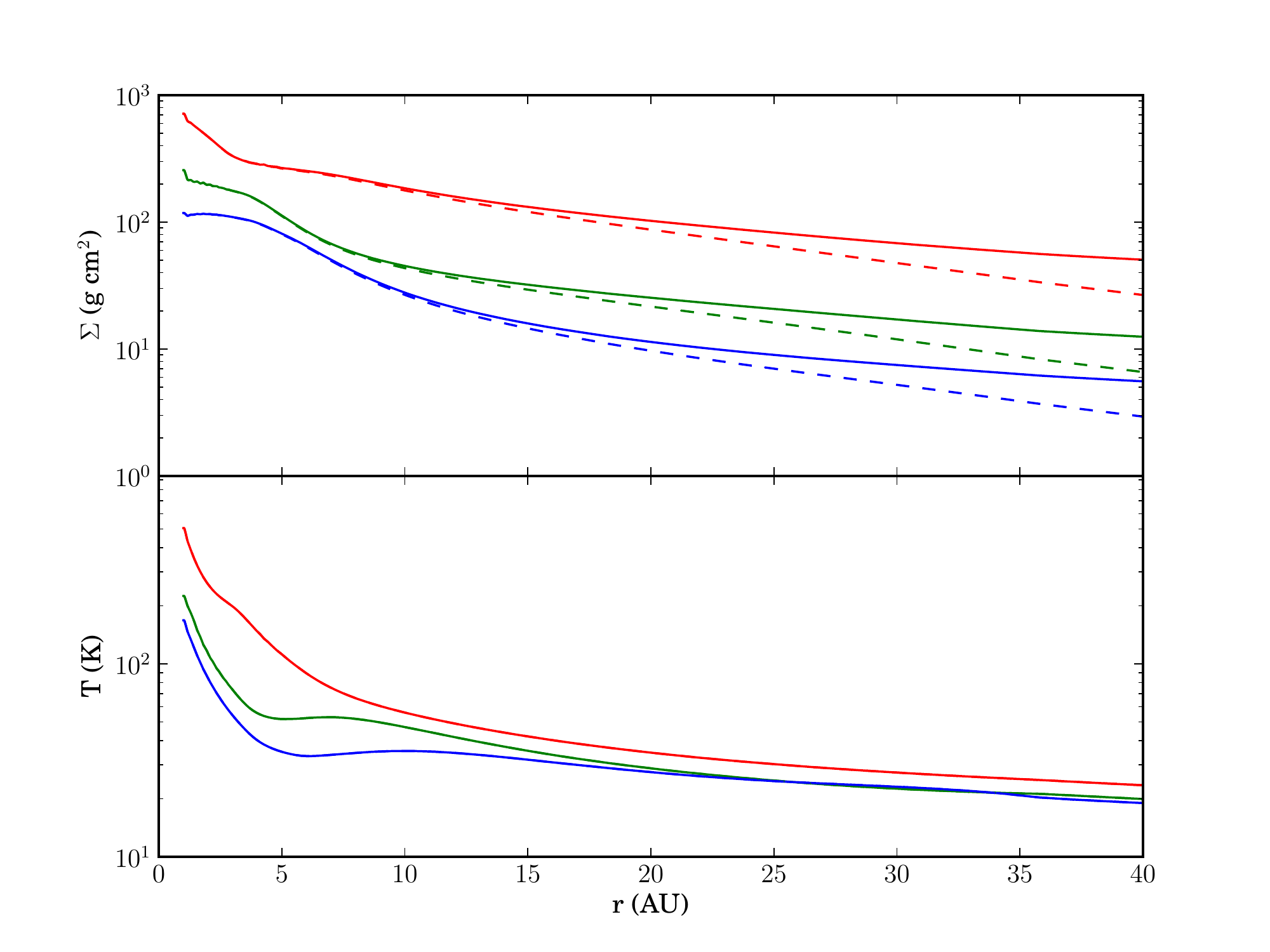}
	\caption{The disk surface density (upper panel) and temperature (lower panel) profiles assumed in this paper.  The disks have mass accretion rates of $5\times 10^{-8}$, $10^{-8}$ (fiducial), and $5\times 10^{-9}$ $M_\odot \rm{yr}^{-1}$ in red, green and blue, respectively.  The dashed-curves indicates how we modify the surface density to account for an exponential fall off as described in Eq.~\ref{eq:cutoff}.
	\label{fig:TSigma}}
\end{figure}	

The green curves in figure~\ref{fig:TSigma} show the surface density and temperature profile for our fiducial disk model.  It has a mass accretion rate of $\dot{M} = 10^{-8} M_\odot {\rm yr}^{-1}$ around a 1 Myr old star.  
The disk has a viscosity $\nu=\alpha c_s^2 \Omega^{-1}$ where $\alpha = 5.4\times 10^{-3}$.
We note that \citet{Bitsch.etal.2015} assume a steady-state accretion profile throughout the disk.  This means that the gas surface density, $\Sigma$, is directly determined by the mass accretion rate and viscosity by $\dot M = 3\pi\nu\Sigma$ \citep{Pringle.1981}. 
The other solid curves in figure~\ref{fig:TSigma} show how the surface density and temperature depend on the mass-accretion rate. As accretion disks are expected to lose mass over time, these higher ($5\times 10^{-8}$ $M_\odot \rm{yr}^{-1}$) and lower ($5\times 10^{-9}$ $M_\odot \rm{yr}^{-1}$) mass accretion rates can be thought of as earlier and later evolutionary stages of the disk, respectively.  

This steady-state assumption is expected for the inner region of any viscously evolving disk \citep{Pringle.1981}.  However, real disks do not extend forever, and as one enters into the outer region of the disk viscous evolution will lead to a smooth decrease in the disk surface density.  Observations of disks suggest that an exponential cut-off is an appropriate approximation for the outer disk surface density profile \citep{Andrews.etal.2010}.  This cutoff will modify the surface density of the disk significantly, but it will only make minor changes to the disk temperature profile.  That is because the thermal structure of the outer region of the protoplanetary disk is dominated by the passive re-radiation of the intercepted stellar light, a process with only very weak dependencies on the disk surface density.
Therefore we approximate the disk parameters in a disk with an exponential cutoff as the dashed curves in figure~\ref{fig:TSigma}, where the surface density is 
\begin{equation}
	\Sigma = \frac{\dot{M}}{3\pi\nu}\exp\left(\frac{-r}{50\,{\rm AU}}\right). 
	\label{eq:cutoff}
\end{equation}
and we assume the temperature profile of this disk is unchanged.

In figure \ref{fig:size} we show the sizes of such pebbles in our fiducial
disk, assuming $\rho_s = 0.5\,{\rm g\,cm}^{-3}$. The solid curves show the size of particles with a fixed $\tau$ (ranging from $10^{-2}$ to 1) in our fiducial disk with $\dot{M}=10^{-8} M_\odot\,\rm{yr}^{-1}$.  This range of $\tau$ is the size range most susceptible to the streaming instability, so we highlight it with gray shading. 
The dashed-curves (and light gray region) shows how the Stokes numbers would vary in the disk if we modify our fiducial disk by assuming an exponential cutoff.  
This figure demonstrates how the size of ``pebbles'' decreases with heliocentric distance in normal protoplanetary disks. 
%The colored boxed regions are described in section \ref{sec:comets}.  
The sizes of those ``pebbles'' is sensitive to the disk properties.
In figure \ref{fig:size2} we highlight the same range of $\tau$ in disks with higher ($5\times 10^{-8}$
$M_\odot \rm{yr}^{-1}$)  and lower ($5\times 10^{-9}$ $M_\odot \rm{yr}^{-1}$) mass accretion rates, corresponds to earlier and later evolutionary stages of the disk, respectively.  From figure~\ref{fig:TSigma} we can see that the largest difference between these disks is the change in the surface density, however, at any given time, larger pebble are expected to be incorporated into comets in the inner disk as compared to the outer disk.

%Because the biggest change for most of the comet formation region is the surface density, not the temperature, as these regions are dominated by stellar irradiation.  Therefore these high and low mass accretion rate models are similar to models in which we decreased or increased the assumed viscosity, respectively.

%When pebbles reach a Stokes number of 1 they are maximally mobile in the disk.
%Therefore, one may posit that pebbles should reach the Stokes number of unity
%(or some fixed fraction of it) before being lost to drift or destructive
%fragmentation.  While larger particles may participate, other loss mechanisms
%(fragmentation, drift) are likely to prevent a significant fraction of material
%from getting to these larger sizes.  And if these growth barriers can be
%overcome, the justification for the GI collapse becomes less compelling.

%\citet{Birnstiel.etal.2012} found that pebbles would grow to a size limited either
%by fragmentation or by the drift rate being faster than the growth rate.  In
%figure \ref{fig:size} we show these two limiting sizes in our fiducial disk.

%In order to discuss how large pebbles are expected to be in protoplanetary
%disks we must assume the structure of the protoplanetary disk.  

\section{``Pebbles'' in Hartley 2} \label{sec:comets}

In order to determine the aerodynamic properties of these pebbles, one must
know three things: the size and density of the pebbles, the properties of the gaseous
protoplanetary disk, and the location in the disk where the comet formed.

\citet{Kelley.etal.2013} calculated the size of the particles in the coma based
on the observations of individual particles.  They found that the maximum
particle radius of 30 cm if one assumes a bright, icy albedo ($A_p=0.67$) and a
maximum radius of 4 m if one assumed a dark, cometary albedo ($A_p=0.049$),
with a steep size distribution of $dn/da \propto a^{-4.7}$ down to the
detection limit.  \citeauthor{Kelley.etal.2013} take the dark solution to be
unlikely for a number of reasons.  First, if the particles are dark then the largest particles in the size distribution are 8 meters in diameter, which likely should have
been resolvable.  Additionally, the dark model implies that the coma consists of 0.6-14\% of the mass of the nucleus, which likely violates the constraints on the amount of mass lost each orbit \citep{Thomas.etal.2013}.  Furthermore, there is substantial amount of water vapor emitted isotropically around the coma, consistent with water sublimation from large particles in the coma.
However, we will include the dark solution for completeness and it
may be considered an extreme upper limit on the particle sizes.  

We note that due to the long estimated lifetimes \citep{Beer.etal.2006} of
these ice balls and relatively high velocity of these particles
\citep{Hermalyn.etal.2013}, they likely did not sublimate enough to
significantly change their size between the time of ejection and detection.
Therefore we assume the observed size distribution is the same as the size
distribution at launch.  Additionally, while the turn-over size is not directly
observed, the presence of a turn over near the detection limit can be inferred
in the data because otherwise unresolved particles would cause a diffuse
emission that would have been detected \citep[see][for
details]{Kelley.etal.2013}.  Therefore in our plots we assume that the
differential particle size distribution is flat between the detection limit
($10^{-12}\,{\rm W\,m^{-2}\,\mu m^{-1}}$) and a factor of two below the
detection limit. This leads to most of the mass of the pebbles having sizes of
either 3 to 5 cm or 40 to 60 cm assuming a bright or dark albedo, respectively.

\begin{figure}[t!]
	\includegraphics[width=\textwidth]{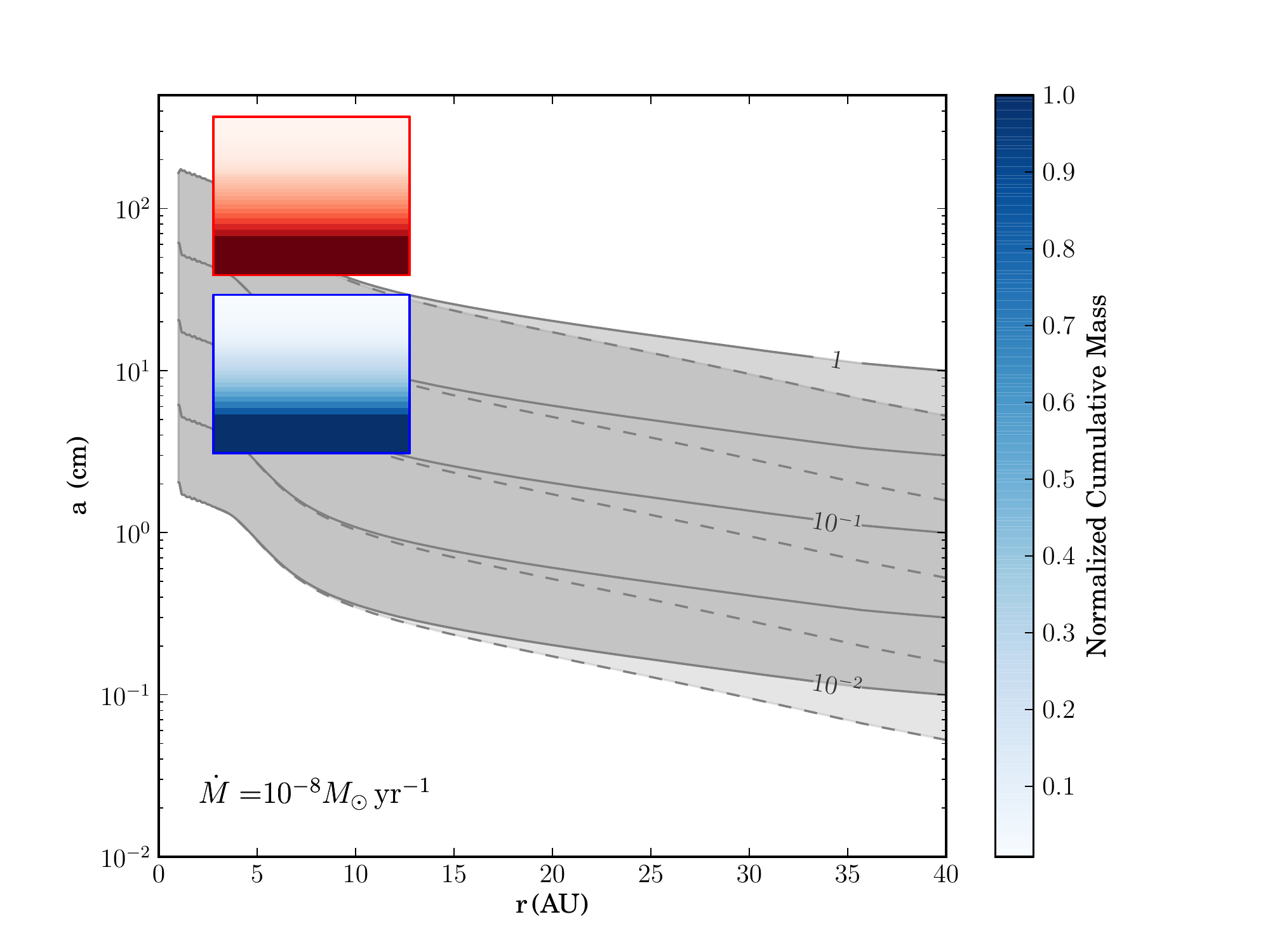}
	\caption{The gray shaded region indicates the size range of particles with density $\rho_s = 0.5\,{\rm g\,cm}^{-3}$ susceptible to concentration via the streaming instability in the nominal disk.  The black curves indicate curves of constant $\tau$, of 1, 0.3, $10^{-1}$, $3\times 10^{-2}$, $10^{-2}$.  The dashed curves indicate how the $\tau$ values would change if we modify the surface density by adding an exponential cutoff as in Eq.~\ref{eq:cutoff}.  The color gradients within the blue and red boxes indicate the range of sizes of particles observed in the coma of comet Hartley 2 if a bright or dark albedo is assumed, respectively.  The shading corresponds to the normalized fraction of the pebble mass in each size bin.  The boxes are placed at a temperature range consistent with the comet's chemistry as described in Sec.~\ref{sec:comets}.
	\label{fig:size}}
\end{figure}	

It is extremely difficult to determine the formation location of any comet.
Comet 103P is a Jupiter family comet (JFC), and from the inclination
distribution of this population it is clear that the most likely immediate
source region of these comets is the Kuiper Belt region \citep[R $>$ 30
AU,][]{Mumma.etal.1993}. However, modern models of solar system evolution
predict that there was significant dynamical mixing early on
\citep{Levison.etal.2008} so that planetesimals from the giant planet forming
region (5-14 AU) were scattered both into the Oort cloud and the scattered
disk, some of which later evolved on to JFC orbits.  Indeed, the varied
composition of JFCs (such as the wide range of D/H ratios
\citet{Hartogh.etal.2011,Altwegg.etal.2015}) supports the idea of mixing.

Fortunately, we have additional constraints on the formation location of this
comet---its composition.  Comet 103P has a D/H ratio of $1.4\times 10^{-4}$,
similar to that of the Earth \citep{Hartogh.etal.2011}.  This make it seem
reasonable that comet 103P formed closer to the sun than most other well
studied comets that have higher D/H ratios \citep{Delsemme.1998}.  Additionally,
comet 103P also has a notably high CO$_2$/CO ratio of around 100:1, compared to
the 1:1 or less more typical of other comets which have a high enough CO$_2$
ratio to be observed, suggesting that it may have formed significantly interior
to the CO snow line.  This is again in contrast to most well-characterized
comets \citep{AHearn.etal.2012}.  We would like to use this large CO$_2$/CO
ratio to give precise constraints on the comet's formation location.
Unfortunately, the exact pathway for CO$_2$ formation in comets is still not
well understood, and different mechanisms give rise to different formation
locations for 103P \citep[e.g.][]{Garrod.Pauly.2011,Noble.etal.2011}.  One
possibility is that the comet formed near or just outside of the CO$_2$ snow
line (at a formation temperature around 42-52K) \citep{Oberg.etal.2011}.
However, CO$_2$ is not prevalent in the gas phase, so direct condensation of
CO$_2$ may not actually be that efficient at producing CO$_2$ ice.
Therefore, another possible pathway for CO$_2$ ice production is by the
conversion of CO trapped in amorphous water ice.  This process has been
demonstrated to be efficient even up to 77 K \citep{Yokochi.etal.2012}.
Therefore, we will take the stance that the high CO$_2$/CO suggests that the
comet most likely formed in a region of the protoplanetary disk with a
temperature between 80 and 40 K.  We note that these temperatures are warmer
than the spin temperature of water measured in Hartley 2
\citep{Kawakita.etal.2004a}, however, that ratio may be modified by the
desorption process, therefore may not be reliable indicator of formation
location \citep{vanDishoeck.etal.2014}.

\begin{figure}
	\includegraphics[width=\textwidth]{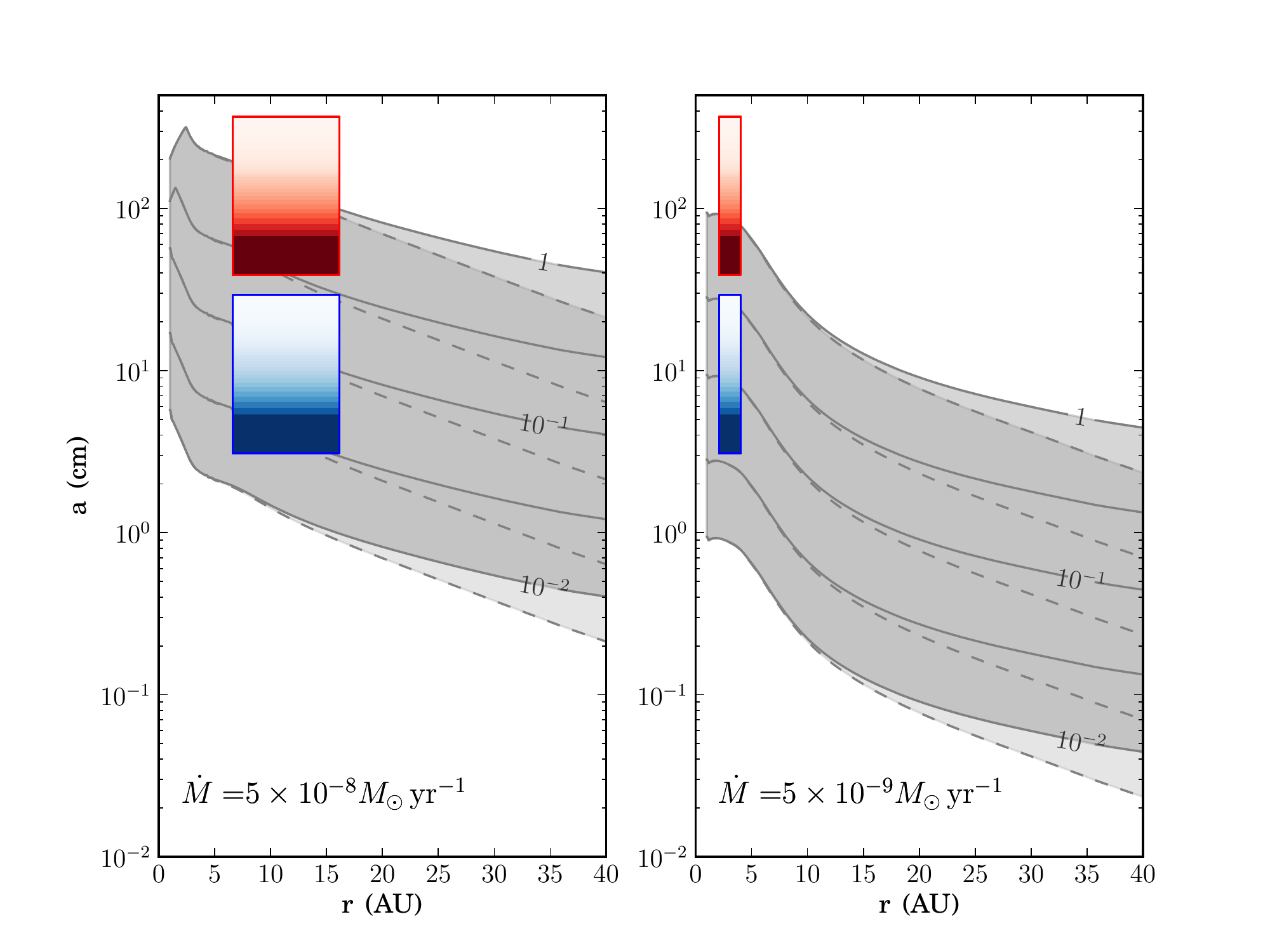}
	\caption{The size range of particles susceptible to concentration via the streaming instability in a high-mass accretion rate on the left, with a low-mass accretion rate on the right.  The curves and gradients are similar to those in Figure~\ref{fig:size}.
	\label{fig:size2}}
\end{figure}

In figure \ref{fig:size} we overlay the size distribution of particles observed
in the coma of Hartley 2 at the range of temperatures we believe this comet
likely formed, between 80 and 40 K.  With the preferred ``bright'' albedo, we
find the particles would be consistent with a $\tau$ range from
$2\times10^{-2}$ to 1, consistent with expectations for the streaming
instability.  If the ``dark'' solution turns out to be correct, then the
largest particles in the coma are would likely have been larger than $\tau=1$, and thus is it not understood how particles of that size could have formed planetesimals.
The majority of the mass is still in the expected range
(particularly if the comet formed relatively close to the sun) so these large
particles could be aggregates of primordial pebbles, but more detailed
consideration would likely be necessary.  While the Stokes number of pebbles is
dependent on the disk properties, for our high and low mass-accretion rate disk
models in figure~\ref{fig:size2} the bright pebbles still agree quite well with
what is needed for the streaming instability.

If we assume that comets that formed farther from the sun have a similar range
in $\tau$ to theses pebbles, then they would have a maximum pebble size of
$\sim$5 cm, but, if the size distribution is similar to that of Hartley 2,
most of the mass would be in smaller, sub-centimeter sizes.  Radar
depolarization has indicated that there must be a significant population of
coma grains larger than 2 cm in size in comet C/2001 A2 (LINEAR)
\citep{Nolan.etal.2006}.  The spin temperature of this comet is colder than
that of Hartley 2 \citep{Kawakita.etal.2004a,DelloRusso.etal.2005}, but
unfortunately there is neither a D/H ratio nor a CO/CO$_2$ ratio to narrow
down the formation location.  In general, most comets do not have signatures
of such large particles.  There are a few exceptions.  For instance
\citet{Reach.etal.2000} found particles as large as 5 cm in the tail of comet
Encke a comet which does have detectable CO$_2$ \citep{Reach.etal.2013}, but
does not have detectable CO, although it is not clear how significant thermal
processing may be obsecuring the chemical composition
\citep{Fernandez.etal.2005}.  Additionally, other detections of large particles
consist of mm or up to cm sizes
\citep[e.g.][]{Harmon.etal.2004,Trigo-Rodriguez.etal.2008,Trigo-Rodriguez.etal.2010}.
This general lack of very large particles appears consistent with the idea that
these comets (which more likely formed in the outer regions of the Solar System
due to the D/H and CO/CO$_2$ ratios) should have been formed from smaller
pebbles.

\section{Conclusions and Discussion}\label{sec:conclusions}

Comet Hartley 2 is observationally a bit of an outlier in two different and
seemingly unrelated facets: it both is the comet with the largest particles
observed in its coma and it has chemistry indicating that it formed at a warmer
location than typical comets.  Given that current theoretical models suggest
that comets may have formed from the accretion of ``pebbles'' with Stokes
numbers of $10^{-2}$ to 1, we postulate that these two observations are not in
fact independent.  The large particles observed in the coma of Hartley 2 are
consistent with this size of pebbles for a range of reasonable disk models
given the likely formation location of the comet based on its chemistry.
Therefore, while we cannot conclusively state that these indeed are primordial
pebbles, it is at least an intriguing possibility that we may be seeing some
remnant of these initial pebbles.  

If this is true then we expect a trend in constituent pebble size commensurate
with their formation location.  Therefore, we expect that the majority of
comets, which have chemistry consistent with formation in the outer regions of
the protoplantary disk, will have smaller pebbles than those seen in Hartley 2. 
We note that much smaller pebbles cannot explain the activity of comets if
their activity is driven purely by water, however as these comets likely have
significant amount of CO this may not prove problematic
\citep{Blum.etal.2014,Blum.etal.2015}.  For example, based on the high D/H
ratio \citep{Altwegg.etal.2015} and the N$_2$/ CO ratio \citep{Rubin.etal.2015}
of comet 67P/Churyumov–Gerasimenko, we may expect that it formed in the
colder-outer regions of the Solar System and thus would have been likely formed
from small, perhaps cm or sub-cm sized pebbles (although we note that the high tensile strength of small particles may require that the pebbles be at least cm in size to explain the observed activity levels \citep{Gundlach.etal.2015}).
Indeed, there previous detections of dust in the coma of 67P suggest that the
tail and coma is dominated by large 100$\mu$m-mm sized particles
\cite{Kelley.etal.2008,Bauer.etal.2012,Rotundi.etal.2015} although there were
also detections of larger particles, in cm to meter sizes
\citep{Rotundi.etal.2015,Sierks.etal.2015}.  Under the arguments presented in
this paper we believe these larger, dusty objects are aggregates, pieces of the
cometary crust crust that have been ejected rather than primordial ``pebbles.''
Similarly, we expect the ``goose bumps'' observed on the surface of 67P
\citep{Sierks.etal.2015}
%, Weissman.AHern.2015} 
are not signatures of their component pebbles but instead have been formed by
some other mechanism modifying the cometary surface.

\section{Acknowledgments}
We would like to thank M.~Mumma, P.~Weissman and M.~Kelley for useful
discussions We would also like to thank J\"urgen Blum and anonymous referee for
their thoughtful suggestions, which have improved this manuscript.  This work
is supported by the Goddard Center for Astrobiology (a NASA Astrobiology
Institute).



%\bibliography{main}


\end{document}